\documentclass[a4paper]{article}

\usepackage{url}
\usepackage{xspace,subfigure}
\usepackage{multirow}
\usepackage{graphicx}

\def\+{\!+\!}
\def\-{\!-\!}

\def\pref(#1,#2){$#1$ is a prefix of $#2$}
\def\suff(#1,#2){$#1$ is a suffix of $#2$}
\def\reg(#1,#2){$#2$ is $#1$-regular}
\def\notreg(#1,#2){$#2$ is not $#1$-regular}

\def\eqref#1{(\ref{#1})}

\newlength{\onedigit}
\newcommand{\w}{\makebox[\onedigit]{~}}
\newlength{\onecomma}
\newcommand{\comrad}{{\sc Comrad}}
\newcommand{\rlz}{{\sc Rlz}}
\newcommand{\repair}{{\sc Re-pair}}
\newcommand{\dnax}{{\sc Dna-x}}

\begin{document}

\title{\bf Reference Sequence Construction for Relative Compression of Genomes
\footnote{
This work was supported by the Australian Research Council and
the NICTA Victorian Research Laboratory.  NICTA is funded
by the Australian Government as represented by the Department
of Broadband, Communications and the Digital Economy and the
Australian Research Council through the ICT Center of Excellence
program. \textit{A version of this paper is to appear in the Proceedings
of SPIRE 2011.}}
}

\author{
Shanika Kuruppu\footnote{National ICT Australia, Department of Computer Science \& Software Engineering, University of Melbourne, Australia, \texttt{\{kuruppu,jz\}@csse.unimelb.edu.au}}
\and
Simon J.~Puglisi\footnote{School of Computer Science and Information Technology, Royal Melbourne Institute of Technology, Australia, \texttt{simon.puglisi@rmit.edu.au}}
\and
Justin Zobel\footnotemark[2]
}

\date{}

\maketitle \thispagestyle{empty}

\begin{abstract}

Relative compression, where a set of similar strings are compressed with
respect to a reference string, is a very effective method of compressing
DNA datasets containing multiple similar sequences. Relative compression
is fast to perform and also supports rapid random access to the
underlying data. The main difficulty of relative compression is in
selecting an appropriate reference sequence. In this paper, we explore
using the dictionary of repeats generated by {\comrad}, {\repair} and
{\dnax} algorithms as reference sequences for relative compression. We
show this technique allows better compression and supports random access
just as well.  The technique also allows more general repetitive
datasets to be compressed using relative compression.

\end{abstract}

\section{Introduction}
\label{sec:Introduction}

Rapid advancements in the field of high-throughput sequencing have
led to a large number of whole genome DNA sequencing projects. Some of these
projects take advantage of the improved sequencing speeds and costs, to 
obtain genomes of species that are unsequenced to date; for
example the Genome 10K project (\url{www.genome10k.org}). Others focus
on resequencing, where individual genomes from a given species are 
sequenced to understand variation between individuals. Examples are the
1000 Genomes project (\url{www.1000genomes.org}) for humans and the 1001
Genomes project (\url{www.1001genomes.org}) for the plant
\textit{Arabidopsis thaliana}. The assembled sequences from these
projects can range from terabytes to petabytes in size. Therefore,
algorithms and data structures to efficiently store, access and search
these large datasets are necessary. Some progress has already been
made~\cite{Brandon2009,Christley2009,Kreft2010,Makinen2010,Kreft2011}, but 
significant challenges remain.

DNA sequences may contain repeated substrings within a sequence,
however, in a database of sequences, the most significant repeats occur
{\em between} sequences, usually those of the same or similar species.
To help manage large genomic databases, compression algorithms that
capture and efficiently encode this repeated information are employed.
Compression algorithms specific to DNA sequences have been around for
some
time~\cite{Behzadi2005,Cao2007,Chen2000,Chen2002,Grumbach1993,Grumbach1994,Rivals1996}.
However, most existing algorithms are unsuitable for compressing large
datasets of multiple sequences.  More recently, algorithms that compress
large repetitive datasets, that also support random access and search on
the compressed sequences, known as \textit{self-indexes}, have emerged.
Some of these algorithms are specific to DNA compression and support
random access queries~\cite{COMRAD2011,RLZspire2010}. Others can
compress general datasets and also implement search queries on the
compressed sequences~\cite{Kreft2010,Makinen2010}.

One of the most effective ways to compress a repetitive dataset
containing multiple sequences from the same or very similar species, or
sequences serving the same biological functions, is to compress each
sequence with respect to a chosen reference
sequence~\cite{Cao2007,RLZspire2010,Makinen2010}. The need for such a
compression method for DNA sequences was first realised by Grumbach and
Tahi~\cite{Grumbach1993}. \texttt{XM}, a statistical algorithm that
implements this feature, can also generate probabilities for the level
of similarity between the reference sequence and the sequence being
compressed~\cite{Cao2007}. Christley, et
al.\ proposed a solution to store just the variations of each human
genome with respect to the reference genome~\cite{Christley2009} and
a similar approach is taken by Brandon, et al.~\cite{Brandon2009}. 
M\"akinen, et al.\ introduce more general methods to compress highly 
repetitive collections which also support searching in the compressed 
data~\cite{Makinen2010}. 

The {\rlz} method, which is used in this paper, represents each 
sequence as an \texttt{LZ77} parsing~\cite{Ziv1977} with respect to a reference
sequence chosen from the dataset~\cite{RLZspire2010}. Recently 
Grabowski and Deorowicz engineered {\rlz} to improve runtime and
compression performance~\cite{Grabowski2011}.

Relative compression algorithms like \texttt{RLZ} produce good
compression results because the reference sequence acts as a static
``dictionary" that includes most of the repeats present in the dataset
being compressed. Compression speed is fast because the sequences can be
compressed in a single pass over the collection, once an index on the
reference sequence has been built. The static reference also makes
random access fast, and easy to support. The main drawback is the
difficulty of selecting an appropriate reference sequence. Selecting a
reference sequence from a dataset containing only individual genomes
from the same strain of the same species is simple, as any sequence will
act as a good reference sequence.  However this will not be effective
for datasets containing sequences from different species, or from
different strains of the same species.

Grabowski and Deorowicz~\cite{Grabowski2011} attempt to address this issue by
adjusting the composition of the reference sequence during
compression. When substrings of a certain minimum length, which do not
occur in the reference sequence, are encountered, they are appended to
the reference sequence, so that later occurrences of those substrings
can be encoded as references. Results in~\cite{Grabowski2011} show that 
such a mechanism can provide a slight improvement to compression 
with no effects on the compression or decompression times. However, this 
method over-compensates and adds more substrings to the reference sequence 
than necessary. We compare our results with those of Grabowski and Deorowicz
in Section~\ref{sec:Results}.

\paragraph{Our contribution:} In this paper we explore the artificial
construction of reference sequences from the phrases built by popular
dictionary compressors.  We find that artifically constructed reference
sequences allow superior compression, while retaining the principle
advatange of relative compression: fast random access to the collection.

\section{Reference Sequence Selection}
\label{sec:DictConstruct}

\begin{figure*}[!t]
\centerline{\includegraphics[scale = 0.3]{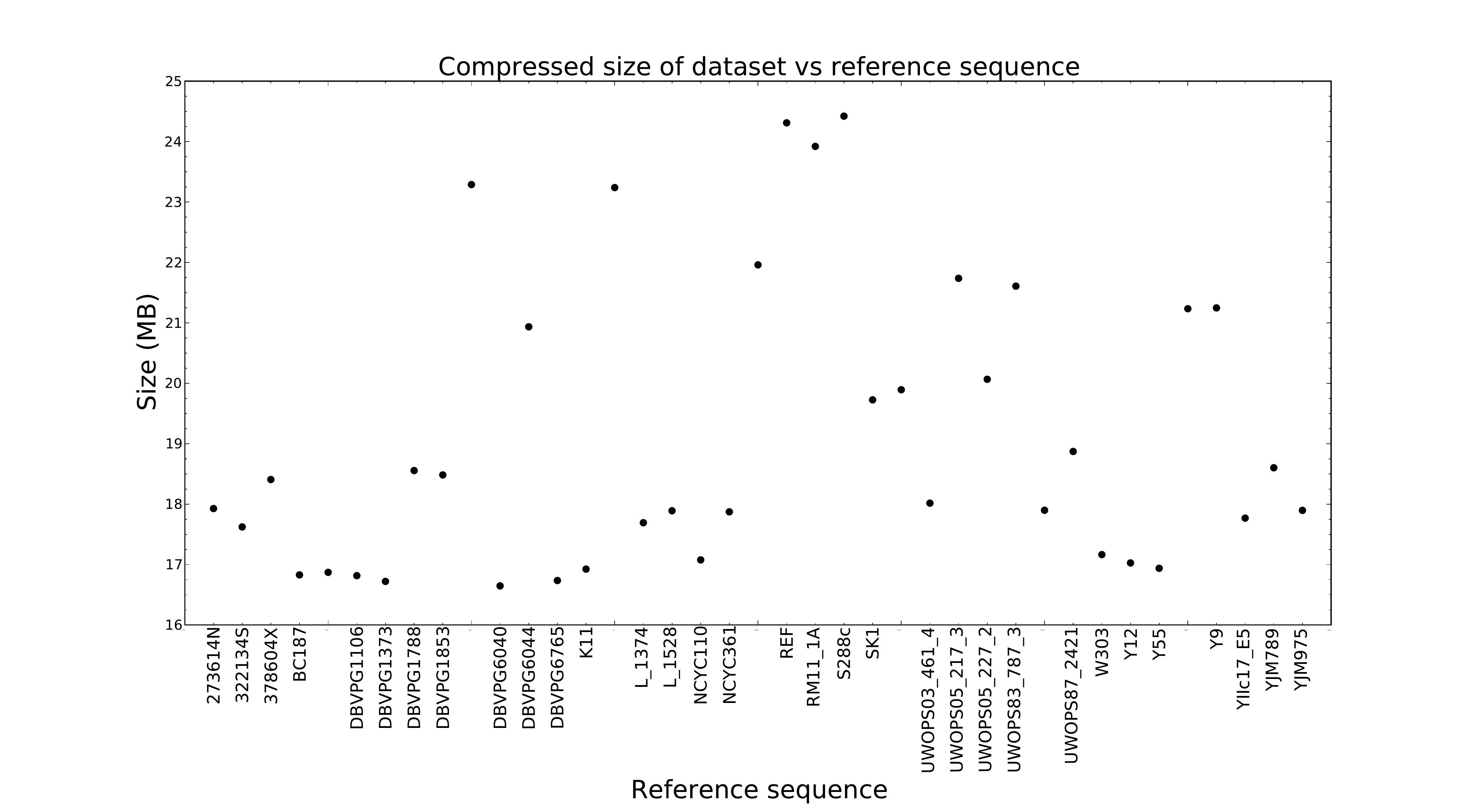}}
\caption{The change in the compressed size of the \textit{S.~cerevisiae}
dataset when the reference sequence is changed. The y-axis contains the
compressed size, measured in Megabytes and the x-axis contains the
reference sequence used.}
\label{fig:Diffref}
\end{figure*}

Before we explore ways to generate an appropriate reference sequence, we
first analyse the effect on compression when ``good" and  ``bad"
reference sequences are used. As an example, we use the {\rlz} algorithm
to compress the \textit{S.~cerevisiae} dataset containing 39 yeast
genomes from different strains. The dataset was compressed 39 times,
with a different sequence being used as a reference each time.
Figure~\ref{fig:Diffref} shows that the reference sequence chosen can
impact compression significantly. For instance,
choosing the sequence \texttt{DBVPG6765} results in a compressed size of
16.65~MB for the \textit{S.~cerevisiae} dataset, while choosing the
sequence \texttt{UWOPS05\_227\_2} results in 
24.42~MB. The experimental results of {\rlz} in \cite{RLZspire2010} uses
the reference genome \texttt{REF} for the \textit{S.~cerevisiae}
species. Using \texttt{REF}, a compressed size of 17.89~MB was achieved,
not far from the best result of 16.65~MB. This example illustrates
that a more principled approach to selecting a reference
sequence is necessary.

The n\"aive way to select the best reference sequence from a dataset is
to follow the approach taken to generate Figure~\ref{fig:Diffref};
compress the dataset many times, each time using a different sequence as
the reference sequence, then select the sequence that gives the best
compression as the reference sequence. In this manner,
\texttt{DBVPG6765} is chosen as the reference sequence for the
\textit{S.~cerevisiae} dataset. This technique is feasible for small
datasets but is ultimately not scalable.

Moreover, a single reference sequence still may not be representative of
the repetitions present in the whole dataset. A sequence may be highly
similar to a few other sequences in the dataset but quite different from
others. In other words, the sequences may form clusters.  This is
plausible for datasets containing genomes from various {\em strains} of
a species. To test this hypothesis, we used the factors that are
generated by {\rlz} that form alignments to the reference sequence (LISS
factors that encode the segments of DNA that are not
mutations~\cite{RLZacsc2011}). We graphed the position component of
these aligning factors for the \textit{S.~cerevisiae} dataset, when
sequence {\tt REF} is used as the reference. If the set of aligning
factors are the same across two sequences, then those two sequences
align to the reference sequence in the same way, hence the two sequences
are similar.

\begin{figure*}[!t]
\centerline{\includegraphics[scale = 0.3]{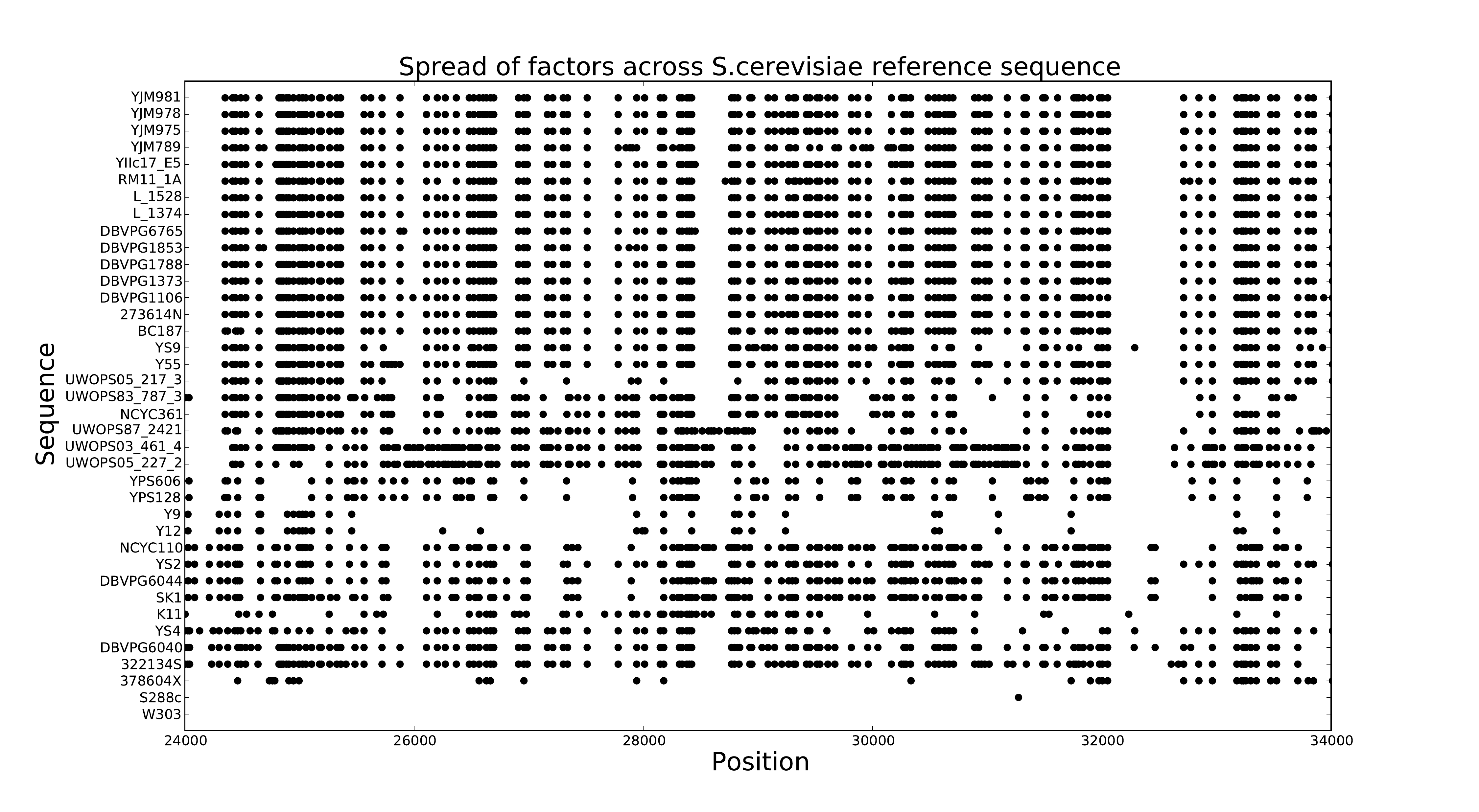}}
\caption{The position components of the first 100 aligning factors for
each sequence in the \textit{S.~cerevisiae} dataset. Only the factors
that start at positions in the range of 24000-34000 are visible. The
x-axis is the position on the reference sequence and the y-axis is the
sequence names for sequences in the dataset.}
\label{fig:LISSfacpos}
\end{figure*}

The aligning factors for each sequence in the \textit{S.~cerevisiae}
dataset for the position range 24,000-34,000 of the reference sequence,
are illustrated in Figure~\ref{fig:LISSfacpos}.  The graph highlights
clusters of similar sequences.  Most sequences have factors that start
at the same position, especially those in the top half of the graph. The
latter half of the graph has clusters of sequences that have similar
factor positions.  As an example, \texttt{YPS606} and \texttt{YPS128}
seem to align to the reference sequence in the same way, and so do the
sequences \texttt{UWOPS03\_461\_4} and \texttt{UWOPS05\_227\_2}.

An alternative to using multiple reference sequences is to use a single
reference sequence that includes the significant repeats in the whole
dataset. The substrings that are shared among the sequences within
clusters can be used to create a reference sequence.  Dictionary
compression algorithms find the repeated substrings of the dataset being
compressed and the dictionary stores these repeats. Hence, a dictionary
compression algorithm that detects global repetitions can be used to
generate a dictionary whose entries can then be concatenated to
construct a reference sequence. We experiment with this idea next.

\section{Reference Sequence Construction}

We choose three dictionary compression algorithms to generate reference
sequences for the two yeast datasets; {\repair}~\cite{Larsson1999}, a
well-known dictionary compression algorithm,
{\comrad}~\cite{COMRAD2011}, similar to {\repair} but tailored for DNA
compression, and {\dnax}~\cite{Manzini2004}, a DNA-specific
implementation of the algorithm by Bentley and
McIlroy~\cite{Bentley1999}.  We first compress our test datasets with
{\repair}, {\comrad} and {\dnax}, and then use the dictionary of repeats
as a reference sequence for relative compression. Below we explain each
algorithm briefly and the process used to generate the reference
sequence from the dictionary.

\subsection*{\repair}
\label{sec:Repair}

The {\repair} algorithm~\cite{Larsson1999} operates in multiple
iterations. In the first iteration, a count of all the distinct pairs of
symbols in the input sequence are recorded. Then the most frequent
symbol pair is replaced by a new symbol, and the counts are updated to
reflect the replacement.  In this manner, the algorithm substitutes the
symbol pair with the highest count at each iteration, until there are no
symbol pairs left with a count of more than one. The new symbols
generated by the algorithm are identified as `non-terminals', while the
symbols in the original input are identified as `terminals'. The
algorithm outputs the input sequence with all its repeated substrings
replaced by non-terminal symbols, and a dictionary of rules that map all
non-terminals to the symbol pairs that they replaced. The dictionary is
hierarchical, since during later iterations, rules of the form $B
\leftarrow CD$ or of the form $B \leftarrow cD$ or $B \leftarrow Cd$ are
generated, where upper-case symbols are non-terminals and lower-case
symbols are terminals.  The non-terminals $C$ and $D$ in turn may also
represent other non-terminals and so on.

The dictionary of rules generated by {\repair} contains the repeated
substrings of the input sequence. The right hand sides of the rules can
be expanded recursively to obtain the repeated substrings, which can
then be concatenated to create a reference sequence.  It's not
necessary to add all of the expanded rules to the reference sequence.
Some of the rules lower in the hierarchy have already been incorporated
into the repeated substrings of rules higher in the hierarchy that refer
to these rules, so it is redundant to add these to the reference
sequence. For example, expanding rule $Z$ in the set of rules $Z
\leftarrow XY$, $X \leftarrow aA$, $Y \leftarrow CD$, would result in
rules $X$, $Y$, $A$, $C$ and $D$ being expanded. Once $Z$ is expanded,
it is redundant to individualy expand $X$, $Y$, $A$, $C$ and $D$. To
implement this, we use a bit vector that is the length of the total
number of rules. To begin with, all the bits are set to zero. When a
rule $Y$ appears on the right hand side of another rule $Z$, then the
bit for rule $Y$ is set to 1 to indicate that when it is $Y$s turn to
get expanded later, it can be skipped.

The non-terminals generated by {\repair} are identified using unique
integers. The higher the non-terminal number, the later the rule was
generated and the higher up in the hierarchy the rule is likely to be.
So starting from the highest numbered rule to the lowest numbered rule,
rule $Z$ is expanded if and only if $Z$ has not been expanded by a
previous rule, as indicated by the bit vector. If rule $Z$ is expanded,
then the resulting substring is appended to the reference sequence.
This continues until all of the rules are considered for expansion. The
concatenation of the expanded substrings forms the reference sequence.

\subsection*{\comrad}
\label{sec:Comrad}

Similar to {\repair}, {\comrad}~\cite{COMRAD2011} is a dictionary
compression algorithm that detects repeated substrings in the input, and
encodes them efficiently to achieve compression.  {\comrad} also
operates in multiple iterations, however, it is a DNA-specific
disk-based algorithm designed to compress large DNA datasets. Instead of
replacing pairs of frequent symbols, {\comrad} replaces repeated
substrings of longer lengths to reduce the number of iterations.

The first iteration of {\comrad} counts distinct $L$ length substrings
and the repeated substrings from most frequent to least frequent are
replaced with non-terminals and a dictionary is formed.  The input
sequence now consists of a combination of terminals and non-terminals.
In subsequent iterations, the counts of distinct substrings that satisfy
a certain set of patterns is recorded (see~\cite{COMRAD2011}), and again
substrings from most frequent to least are replaced with non-terminals.
The iterations continue until there are no substrings of the above form
remaining with at least a count of $F$ (only substrings with frequency
$F$ are eligible for replacement). The algorithm outputs the input
sequence with repeated substrings replaced by non-terminals, and like
{\repair}, a dictionary containing the non-terminals mapping to the
substrings they replace.  As with the {\repair} dictionary, we expand
non-terminals and append them to create a reference sequence.

\subsection*{\dnax}
\label{sec:dnaX}

Unlike {\repair} and {\comrad}, {\dnax} is a single pass dictionary
compression algorithm. As the input is read, the fingerprint of every
$B$-th substring of length $B$ is stored in a hash table. To encode the
next substring, all overlapping $B$-mers in the so far unencoded part of
the input are searched for in the hash table until there is a match. The
hash table gives the positions of the earlier occurrences of the
$B$-mer. Each of these occurrences is checked to find the longest
possible match. Then the prefix until the matching substring, followed
by the reference for the matching substring is encoded. Searching and
encoding continues until no more symbols remain to be encoded. The
longest matching substrings encoded by the algorithm are the repeated
substrings we use to construct the reference sequence.  We
modified the implementation of {\dnax} by Manzini and Rastero to only
output the concatenation of the longest matching substrings detected by
the algorithm. We use this output as the reference sequence.

\section{Experimental Results}
\label{sec:Results}

To test the performance of the reference construction method, we use
{\rlz} as the relative compressor. We use three test datasets containing
repetitive genomes: 39 strains of \textit{S.~cerevisiae} and 36 strains
of the \textit{S.~paradoxus} species of yeast, and 33 strains of
\textit{E.~coli} bacteria.  We ran {\repair}, {\comrad} and {\dnax} on
all three datasets. For {\repair}, we used the default parameters, which
does not place any restrictions on the number or length of repeats that
can be detected.  For {\comrad}, we used a starting substring length $L$
of 16 and a threshold frequency $F$ of 2.  For {\dnax} we set the
substring length $B$ to 16 to be consistent with {\comrad}.  The
repeated substrings resulting from the dictionaries were used to
generate the reference sequence as described above. 

Compression results are in Table~\ref{tab:DictCompressRefStd}. The first
section contains the results for compressing with {\rlz} using the
original reference sequence.  The number of megabases (including the
reference sequence) and the 0-order entropy of the dataset are in the
first row.  The second and third row contains the compression results
from using the reference sequences available in the dataset with the
\texttt{RLZ-std} and \texttt{RLZ-opt} (with the full set of
optimisations), respectively.  The results show that \texttt{RLZ-opt}
achieves better compression compared to \texttt{RLZ-std}.

The second section of Table~\ref{tab:DictCompressRefStd} contains
results for using the {\comrad} generated reference sequence.  The two
rows contain results for using the standard implementation of {\rlz}
(\texttt{RLZ-std-C}) and the optimised {\rlz} with look-ahead and short
factor encoding enabled (\texttt{RLZ-opt-C})\footnote{LISS factor
encoding was not used as the reference is not a sequence from the
dataset and so there is no reason to expect factor positions to be
predictable.  For completeness, we compressed with the LISS option on
and the compression results were worse than standard {\rlz}.},
respectively.  The \textit{S.~cerevisiae} and \textit{S.~paradoxus}
datasets compress better using the {\comrad} generated reference
sequence.  The biggest improvement (a factor of two) is for
\textit{E.~coli}.  The original reference sequence was the \texttt{K12}
strain from the dataset, since the species does not have a reference
genome.  Evidently \texttt{K12} is not a sequence that represents the
dataset well and the {\comrad} generated reference sequence is a much
better representation.

The third section of Table~\ref{tab:DictCompressRefStd} contains the
results for the {\repair} generated reference sequences, which are very
similar to the {\comrad} results. The compression results improved for
all three datasets with the most significant improvements being for
\textit{E.~coli}. Overall, using the {\repair} generated reference
sequences led to slightly better compressed sizes than using the
{\comrad} generated reference sequences.

The {\dnax} generated reference sequences are not as promising.  We
found {\dnax} generated large reference sequences, as some of the
repeats it output were redundant.  For example, the reference sequences
for \textit{S.~cerevisiae} are 124.46~Mbases, 127.95~Mbases and
439.27~Mbases for {\repair}, {\comrad} and {\dnax}, respectively.
Filtering such duplicate repeats is difficult as there are no
non-terminal numbers to identify multiple occurrences of the same
repeat. 

Next we show that using a reference sequence containing repeats from the
whole dataset is better than using a single sequence from the dataset as
a reference. As in Section~\ref{sec:DictConstruct}, for all three
datasets, we ran {\rlz}-opt multiple times, with each sequence from the
dataset being used as a reference at each iteration, to select a single
sequence from each dataset that achieves the best compression result
when used as a reference. The best compression results achieved were
9.33~MB, 13.23~MB and 18.69~MB for \textit{S.~cerevisiae} using the
reference genome, \textit{S.~paradoxus} using the \texttt{Z1} strain and
\textit{E.~coli} using the \textit{Sakai} strain, respectively.
Comparing these results to those in the second and third sections of
Table~\ref{tab:DictCompressRefStd} shows that even if the sequence that
gives the best compressed size is chosen as the reference sequence for a
dataset, the compression results are still worse than the results that
could be achieved by using a {\comrad} or {\repair} generated reference
sequence. This confirms that a single sequence is unlikely to capture
all the repeats in a dataset of similar sequences and it is worth
constructing a reference sequence that captures all the significant
repeats of the dataset to achieve better compression results.

Table~\ref{tab:DictCompressRefStdTime} shows compression and
decompression times. Obviously the compression time increases
significantly when using a generated reference sequence as the reference
must now be generated.  Also generated references tend to be longer and
so more time is needed to construct suffix and LCP arrays used to
perform the {\rlz} parsing, and to compress the reference sequence with
\texttt{7zip}. This is particularly the case for {\dnax}.  Still,
performance for all methods remains at an acceptable level: the two
largest datasets, can be compressed in approximately 20 minutes.  More
importantly, decompression times are not affected at all. 

Table~\ref{tab:OtherAlgResults} shows compression results for {\sc
Rlcsa}, {\sc Lz-end}, {\comrad}, {\tt XM} and {\repair} algorithms being
used to compress the three test datasets. The results clearly show that
using {\rlz} with the {\comrad} or {\repair} generated dictionaries
achieve much better compression than even the best results in
Table~\ref{tab:OtherAlgResults}.

While {\repair} generated reference sequences seem to compress the
datasets a little better than those of {\comrad}, resource requirements
of the algorithms should be taken into account. Both {\comrad} and
{\repair} have comparable runtimes ({\repair} required a little over
half the time of {\comrad}, see Table~\ref{tab:DictCompressRefStdTime}).
However, the main memory usage of {\repair} is much higher, with
\textit{S.~cerevisiae} and \textit{S.~paradoxus} using approximatly
12~Gb and 11~Gb, respectively.  On the other hand, {\comrad} only
requires 277~Mb and 554~Mb for \textit{S.~cerevisiae} and
\textit{S.~paradoxus}, respectively.  {\dnax} has the lowest resource
usage, but a better process needs to be followed to extract the
necessary repeats from the dictionary to get a better quality reference
sequence.

\begin{table*}[t]
\begin{center} {\small\begin{tabular}{@{} lc c c c c c c c c @{}}
\hline
Dataset     & \multicolumn{2}{c}{S.~cerevisiae} && \multicolumn{2}{c}{S.~paradoxus} && \multicolumn{2}{c}{E.~coli} \\
               \cline{2-3}                         \cline{5-6}                         \cline{8-9}
            & Size         & Ent.   && Size         & Ent.  && Size         & Ent. \\
            & (Mbytes)     & (bpb)  && (Mbytes)     & (bpb) && (Mbytes)     & (bpb)\\
\hline
Original    & 485.87       & 2.18   && 429.27       & 2.12  && 164.90       & 2.00 \\
{\rlz}-std  & {\w}17.89    & 0.29   && {\w}23.38    & 0.44  && {\w}24.27    & 1.18 \\
{\rlz}-opt  & {\w\w}9.33   & 0.15   && {\w}13.44    & 0.25  && {\w}19.30    & 0.94 \\
\hline
{\rlz}-std-C& {\w\w}8.20   & 0.14   && {\w\w}9.64   & 0.18  && {\w\w}8.70   & 0.42 \\
{\rlz}-opt-C& {\w\w}7.99   & 0.13   && {\w\w}9.08   & 0.17  && {\w\w}8.07   & 0.39 \\
\hline
{\rlz}-std-R& {\w\w}7.78   & 0.13   && {\w\w}9.10   & 0.17  && {\w\w}8.21   & 0.40 \\
{\rlz}-opt-R& {\w\w}7.64   & 0.13   && {\w\w}8.67   & 0.16  && {\w\w}7.72   & 0.37 \\
\hline
{\rlz}-std-D& {\w\w}9.80   & 0.16   && {\w}13.38    & 0.25  && {\w}11.06    & 0.54 \\
{\rlz}-opt-D& {\w\w}9.64   & 0.16   && {\w}13.01    & 0.24  && {\w}10.57    & 0.51 \\
\hline
\end{tabular}}
\end{center}

\caption{Compression results for using {\comrad}, {\repair} and {\dnax}
generated reference sequences. The columns are, the identifiers for
{\rlz} version used and algorithm used to generate the reference
sequence, compressed size of the dataset in Megabytes (original dataset
size in Megabases) and average number of bits used per base when
compressed, respectively.  The sections are for compression results of
{\rlz}  when using, {\comrad}, {\repair} and {\dnax} generated reference
sequences, respectively. In the first section, \texttt{RLZ-opt} includes
all the optimisations. In the last two sections, \texttt{RLZ-opt} only
includes looking ahead and short factor encoding.} 

\label{tab:DictCompressRefStd}
\end{table*}

\begin{table*}[t]
\begin{center} {\small\begin{tabular}{@{} lc c c c c c c c @{}}
\hline
Dataset     & \multicolumn{2}{c}{S.~cerevisiae} && \multicolumn{2}{c}{S.~paradoxus} && \multicolumn{2}{c}{E.~coli} \\
               \cline{2-3}                        \cline{5-6}                        \cline{8-9}
            & Comp.     & Dec.  && Comp.     & Dec.  && Comp.   & Dec. \\
            & (sec)     & (sec) && (sec)     & (sec) && (sec)   & (sec) \\
\hline
RLZ-std     & {\w}143   & 9     && {\w}182   & 6     && 125     & 3 \\
RLZ-opt     & {\w}233   & 8     && {\w}241   & 6     && 140     & 3 \\
\hline
RLZ-std-C   & 1561      & 4     && 1619      & 4     && 588     & 2 \\
RLZ-opt-C   & 1783      & 4     && 1832      & 3     && 658     & 2 \\
\hline
RLZ-std-R   & 1170      & 4     && 1134      & 4     && 455     & 2 \\
RLZ-opt-R   & 1482      & 4     && 1353      & 4     && 499     & 2 \\
\hline
RLZ-std-D   & 2272      & 8     && 1787      & 7     && 618     & 4 \\
RLZ-opt-D   & 2901      & 7     && 2492      & 7     && 843     & 4 \\
\hline
\end{tabular}}
\end{center}

\caption{Compression and decompression times for {\comrad}, {\repair}
and {\dnax} generated reference sequences. The columns are: the
algorithm used and the time taken to compress and decompress measured in
seconds, respectively.  Compression times include the time taken to
generate the reference sequences, where necessary.} 

\label{tab:DictCompressRefStdTime}
\end{table*}

\begin{table*}
\begin{center} {\small\begin{tabular}{@{} lc c c c c c @{}}
\hline
Dataset         & \multicolumn{2}{c}{S.~cerevisiae} & \multicolumn{2}{c}{S.~paradoxus}  & \multicolumn{2}{c}{E.~coli} \\
                  \cline{2-3}                         \cline{4-5}                         \cline{6-7}
                & Size          & Ent.      & Size          & Ent.      & Size      & Ent. \\
                & (Mbytes)      & (bpb)     & (Mbytes)      & (bpb)     & (Mbytes)  & (bpb)\\
\hline
Original        & 485.87        & 2.18      & 429.27        & 2.12      & 164.90    & 2.00 \\
{\sc Rlcsa}     & {\w}41.39     & 0.57      & {\w}47.35     & 0.88      & {\w}34.94 & 1.67 \\
{\sc Lz-end}    & {\w}42.52     & 0.70      & {\w}57.18     & 1.07      & {\w}55.25 & 2.68 \\
{\comrad}       & {\w}15.29     & 0.25      & {\w}18.33     & 0.34      & {\w}13.44 & 0.65 \\
{\tt XM}        & {\w}74.53     & 1.26      & {\w}13.17     & 0.25      & {\w\w}8.82& 0.43 \\
{\repair}       & {\w\w}8.85    & 0.15      & {\w}11.75     & 0.22      & {\w}11.89 & 0.58 \\
\hline
\end{tabular}}
\end{center}

\caption{Compression results for the yeast and \textit{E.~coli} datasets
using other compression algorithms. The first row is the original size
for all datasets (size in megabases), the remaining rows are the
compression performance of {\sc Rlcsa}, {\sc Lz-end}, {\comrad},
\texttt{XM} and {\repair} algorithms. The two columns per dataset show
the size in Mbytes and the 0-order entropy (in bits per base).}

\label{tab:OtherAlgResults}
\end{table*}

We next experiment with data sets which do not contain a specific
reference.  These were a \textit{Hemoglobin} dataset containing 15,199
DNA sequences of proteins that are associated with Hemoglobin, an
\textit{Influenza} dataset containing 78,041 sequences of various
strains of the Influenza virus and a \textit{Mitochondria} dataset
containing 1,521 mitochondrial DNA sequences from various species.
Reference sequences were generated for the datasets using {\comrad},
{\repair} and {\dnax}. The results are presented in
Table~\ref{tab:DictCompressRefOther}.

\begin{table*}[t]
\begin{center} {\small\begin{tabular}{@{} lc c c c c c c c c @{}}
\hline
Dataset     & \multicolumn{2}{c}{Hemoglobin} && \multicolumn{2}{c}{Influenza} && \multicolumn{2}{c}{Mitochondria} \\
              \cline{2-3}                       \cline{5-6}                      \cline{8-9}
            & Size      & Ent.  && Size         & Ent.  && Size      & Ent. \\
            & (Mbytes)  & (bpb) && (Mbytes)     & (bpb) && (Mbytes)  & (bpb)\\
\hline
Original    & {\w}7.38  & 2.07  && 112.64       & 1.97  && 25.26     & 1.95 \\
RLZ-std     & {\w}3.81  & 4.13  && {\w}43.65    & 3.10  && {\w}9.31  & 2.95 \\
\hline
RLZ-std-C   & {\w}1.31  & 1.42  && {\w\w}3.31   & 0.23  && {\w}6.55  & 2.07 \\
RLZ-opt-C   & {\w}1.17  & 1.27  && {\w\w}3.00   & 0.21  && {\w}6.05  & 1.92 \\
\hline
RLZ-std-R   & {\w}1.32  & 1.43  && {\w\w}3.00   & 0.21  && {\w}6.69  & 2.12 \\
RLZ-opt-R   & {\w}1.19  & 1.28  && {\w\w}2.82   & 0.20  && {\w}6.20  & 1.96 \\
\hline
RLZ-std-D   & {\w}1.42  & 1.54  && {\w\w}3.68   & 0.26  && {\w}7.13  & 2.26 \\
RLZ-opt-D   & {\w}1.27  & 1.38  && {\w\w}3.49   & 0.25  && {\w}6.59  & 2.09 \\
\hline
\end{tabular}}
\end{center}

\caption{ Compression results for {\comrad}, {\repair} and {\dnax}
generated reference sequences for compressing repetitive datasets that
do not have an explicit reference sequence.  The columns are,
respectively, algorithm used, compressed size in Megabytes (original
dataset sizes in Megabases) and average number of bits used per base.
The sections are results for using {\rlz} with, the first sequence from
each dataset as a reference sequence, and {\comrad}, {\rlz} and {\dnax}
generated reference sequences, respectively.  {\rlz} is run in standard
mode, \texttt{RLZ-std} and optimised mode, \texttt{RLZ-opt}.  }

\label{tab:DictCompressRefOther}
\end{table*}

\begin{table*}[t]
\begin{center} {\small\begin{tabular}{@{} lc c c c c c c c c @{}}
\hline
Dataset     & \multicolumn{2}{c}{Hemoglobin} && \multicolumn{2}{c}{Influenza} && \multicolumn{2}{c}{Mitochondria} \\
               \cline{2-3}                      \cline{5-6}                      \cline{8-9}
            & Comp. & Dec.  && Comp.    & Dec.  && Comp.    & Dec. \\
            & (sec) & (sec) && (sec)    & (sec) && (sec)    & (sec)\\ 
\hline
RLZ-std     & {\w}5 & 1     && {\w}67   & 11    && {\w}11   & 1 \\
\hline
RLZ-std-C   & 20    & 1     && 196      & {\w}1 && 106      & 1 \\
RLZ-opt-C   & 21    & 1     && 235      & {\w}1 && 107      & 1 \\
\hline
RLZ-std-R   & 11    & 1     && 189      & {\w}1 && {\w}64   & 1 \\
RLZ-opt-R   & 12    & 1     && 220      & {\w}1 && {\w}64   & 1 \\
\hline
RLZ-std-D   & 14    & 1     && 247      & {\w}3 && {\w}69   & 1 \\
RLZ-opt-D   & 17    & 1     && 364      & {\w}3 && {\w}72   & 1 \\
\hline
\end{tabular}}
\end{center}

\caption{ Compression and decompression times for {\comrad}, {\repair}
and {\dnax} generated reference sequences are used for compressing
repetitive datasets that do not have an explicit reference sequence. The
columns are respectively, algorithm used, total time taken to compress
and decompress, measured in seconds.  All compression times include time
taken to generate the reference sequence.  }

\label{tab:DictCompressRefOtherTime}
\end{table*}

The first section of Table~\ref{tab:DictCompressRefOther} contains the
performance of {\rlz} when the first sequence in the dataset is chosen
to be the reference.  We only used standard {\rlz}, since the reference
sequences chosen were arbitrary so none of the {\rlz} optimisations will
be an advantage to the compression. The compression results for {\rlz}
are worse than on previous datasets where a specific reference is
available.

The results in the second section of the table are for using {\comrad}
generated reference sequences.  Compression clearly improves for all
three datasets.  The most significant improvement is for the
\textit{Influenza} dataset, followed by the \textit{Hemoglobin} dataset.
The \textit{Mitochondria} dataset did not compress very well but
compression still improves.

Compression also improved significantly for all datasets by using a
{\repair} generated reference. The \textit{Influenza} dataset had the
most significant improvement, followed by \textit{Hemoglobin}. The
\textit{Mitochondria} dataset still does not compress well. The fourth
section of the table contains the results of using the {\dnax} generated
reference. The results have improved compared to using the original
reference sequence, but gains are less than with the other two
algorithms. 

According to Table~\ref{tab:DictCompressRefOther}, if there is enough
repetitions in the dataset, it is feasible to generate a reference
sequence using either {\repair} or {\comrad}, or any other dictionary
compression algorithm, that can be used by {\rlz} to compress any
arbitrary repetitive dataset. There is no significant difference between
using a {\comrad} generated reference sequence over a {\repair}
generated one, however current implementations of {\repair} are less
scalable than {\comrad}. Table~\ref{tab:DictCompressRefOtherTime} shows
the compression and decompression times.

Finally, we compare the new results for \textit{S.~cerevisiae} and
\textit{S.~paradoxus} to those obtained by Grabowski and
Deorowicz~\cite{Grabowski2011}. The results they achieve without the
improved reference sequence are 7.18~Mbytes and 9.62~Mbytes, and with
the improved reference sequence are 6.94~Mbytes and 9.01~Mbytes for
\textit{S.~cerevisiae} and \textit{S.~paradoxus}, respectively. Our best
results are 7.64~Mbyte for \textit{S.~cerevisiae} and 8.67~Mbyte for
\textit{S.~paradoxus}, using {\repair}, which are comparable. It may
be possible to combine the techniques to acheive even better results.

\section{Concluding Remarks}
\label{sec:Conclusion}

Relative compression is a powerful technique for compressing collections
of related genomes, which are now becoming commonplace. In this paper we
have shown that these genomic collections can contain clusters of sequences
which are more highly related than others. We have also shown that impressive
gains in compression can be acheived by exploiting these clusters. Our specific
approach has been to detect repetitions across the dataset and build an
artificial ``reference sequence'', relative to which the sequence is subsequently
compressed. This method retains the principle advantage of relative compression:
fast random access. The drawback is slower compression time, as time must now
be spent finding repeats with which to generate the reference. Future work 
will attempt to address this problem. We also believe it may be fruitful to 
apply clustering algorithms to related genomes to isolate strains.

\bibliographystyle{abbrv}
\bibliography{DictRelativeCompress}

\end{document}